\documentclass{article}
\usepackage{spconf,amsmath,graphicx}
\usepackage{tabularx}
\usepackage{color}
\usepackage{booktabs}
\usepackage{multirow}
\usepackage{tikz}
\usepackage{cite}
\usetikzlibrary{calc,fit,positioning, backgrounds}
\usetikzlibrary{external}
\usepackage{microtype}
\title{Automatic and Perceptual Discrimination between \\ Dysarthria, Apraxia of Speech, and Neurotypical Speech}
\name{Ina Kodrasi$^{*}$, Michaela Pernon$^{\dagger, \ddagger}$, Marina Laganaro$^{\S}$, Herv\'{e} Bourlard$^{*}$\thanks{This work was supported by the Swiss National Science Foundation project no CRSII5\_173711 on ``Motor Speech Disorders: characterizing phonetic speech planning and motor speech programming/execution and their impairments''.
}}
\address{$^*$Speech and Audio Processing Group, Idiap Research Institute, Switzerland\\
  $^{\dagger}$Department of Clinical Neurosciences, Geneva University Hospital, Switzerland \\
  $^{\ddagger}$Laboratoire de Phon{\'e}tique et Phonologie, UMR 7018, CNRS-Universit{\'e} Sorbonne Nouvelle, France\\
  $^{\S}$Faculty of Psychology and Educational Sciences, University of Geneva, Switzerland\\
}
\begin{document}
\ninept
\maketitle
\begin{abstract}
  Automatic techniques in the context of motor speech disorders~(MSDs) are typically two-class techniques aiming to discriminate between dysarthria and neurotypical speech or between dysarthria and apraxia of speech~(AoS).
  Further, although such techniques are proposed to support the perceptual assessment of clinicians, the automatic and perceptual classification accuracy has never been compared.
  In this paper, we investigate a three-class automatic technique and a set of handcrafted features for the discrimination of dysarthria, AoS and neurotypical speech.
  Instead of following the commonly used One-versus-One or One-versus-Rest approaches for multi-class classification, a hierarchical approach is proposed.
  Further, a perceptual study is conducted where speech and language pathologists are asked to listen to recordings of dysarthria, AoS, and neurotypical speech and decide which class the recordings belong to.
  The proposed automatic technique is evaluated on the same recordings and the automatic and perceptual classification performance are compared.
  The presented results show that the hierarchical classification approach yields a higher classification accuracy than baseline One-versus-One and One-versus-Rest approaches.
  Further, the presented results show that the automatic approach yields a higher classification accuracy than the perceptual assessment of speech and language pathologists, demonstrating the potential advantages of integrating automatic tools in clinical practice.
\end{abstract}
\section{Introduction}
\label{sec:intro}
Various conditions of brain damage may disrupt the speech production mechanism, resulting in motor speech disorders (MSDs) that encapsulate altered speech production in different dimensions.
Two primary categories of MSDs are dysarthria and apraxia of speech~(AoS).
Although dysarthria and AoS arise from disruptions at different levels of the speech production mechanism, they manifest through overlapping clinical-perceptual characteristics such as articulation deficiencies, vowel distortions, reduced loudness variation, hypernasality, or syllabification~\cite{Duffy_book_2003, Ziegler_JSLHR_2012}.
Diagnosing the presence of an MSD (i.e., discriminating between neurotypical and impaired speech) is crucial in clinical practice, since the presence of an MSD can be one of the earliest signs of several neurodegenerative disorders~\cite{Rusz_JASA_2013, Rong_2015, tracy_bi_2020}.
Further, an accurate differential diagnosis of the MSD (i.e., discriminating between dysarthria and AoS) is also important, since it can provide clues about the underlying neuropathology~\cite{Duffy_book_2000, Duffy_2008}.
However, because of the difficulty of detecting clinical-perceptual characteristics by ear (particularly in the presence of mild impairments) and because the clinical-perceptual characteristics of dysarthria and AoS overlap, discriminating between dysarthria, AoS, and neurotypical speech is hard for non-experts and even expert inter-rater agreement can be low~\cite{Fonville_JN_2008,Bunton_JSLHR_2008,Haley_AJSLP_2017}.

To complement the perceptual assessment of clinicians, automatic techniques based on pattern recognition models have been proposed.
Typical automatic techniques operate on acoustic features which are handcrafted to reflect impaired speech dimensions.
Many acoustic features have been successfully exploited to characterize impacted phonation and articulation, e.g., fundamental and formant frequencies, jitter, shimmer, Mel frequency cepstral coefficients, or temporal and spectral sparsity~\cite{Tsanas_ITBE_2012, Orozco-Arroyave_Interspeech_2015, Hemmerling_Interspeech_2016, Sapir_JSLHR_2010, Kodrasi_ICASSP_2019, Kodrasi_ITASLP_2019a, Kodrasi_Interspeech_2020}.
In an attempt to capture many impaired speech dimensions, also large-scale feature sets such as openSMILE have been used~\cite{openSMILE,Bocklet_Interspeech_2013,Norel_Interspeech_2018}.
The extracted features are then used to train classifiers such as Support Vector Machines~(SVMs) or Hidden Markov Models~(HMMs).

The majority of state-of-the-art contributions deal with impaired speech arising due to dysarthrias or laryngeal disorders, with AoS being considered only in~\cite{Kodrasi_Interspeech_2020}.
These contributions propose two-class techniques aiming to discriminate between dysarthria and neurotypical speech, laryngeal disorders and neurotypical speech, or dysarthria and AoS.
Three-class techniques aiming to discriminate between dysarthria, AoS, and neurotypical speech have not been considered in the state-of-the-art literature.
Multi-class techniques have seldom been proposed only in the context of laryngeal disorders~\cite{Vaiciukynas_SC_2012,Behroozmand_SPIT_2005, Dibazar_IEMBS_2006}.
In~\cite{Vaiciukynas_SC_2012}, three-class classification of nodular lesions, diffuse lesions, and neurotypical speech is achieved through multiple SVMs in One-versus-One~(OvO) and One-versus-Rest~(OvR) classification approaches.
In~\cite{Behroozmand_SPIT_2005}, three-class classification of edema, nodules, and polyp is achieved through multiple SVMs in an OvO classification approach. 
In~\cite{Dibazar_IEMBS_2006}, five-class classification of laryngeal disorders is achieved through multiple HMMs in an OvR classification approach.

In this paper, we propose a three-class automatic technique for the discrimination of dysarthria, AoS, and neurotypical speech.
Instead of following an OvO or OvR classification approach, we propose to follow a hierarchical classification approach with two SVMs~\cite{Kumar_PAA_2002, Chen_IGRSS_2004}.
The first SVM discriminates between impaired and neurotypical speech whereas the second SVM discriminates between dysarthria and AoS.
To characterize the different impaired speech dimensions, a \mbox{$28$--dimensional} feature vector is constructed.
Since the discriminative power of different features is expected to be different for different groups of speakers, two feature selection blocks are incorporated prior to the two SVMs.

To the best of our knowledge, although automatic techniques are proposed with the primary objective of complementing the perceptual assessment of clinicians, the automatic and perceptual classification accuracy have never been compared in the literature.
In this paper, we also compare the classification accuracy of the proposed automatic technique to the classification accuracy achieved by speech and language pathologists~(SLPs).
A perceptual study is conducted where $20$ SLPs are asked to listen to recordings of dysarthria, AoS, and neurotypical speech and decide which class the recordings belong to.
The proposed automatic technique is evaluated on the same recordings and the automatic and perceptual classification performance are extensively compared.

The presented results on a French database of dysarthria, AoS, and neurotypical speech illustrate the advantages of the hierarchical classification approach in comparison to OvO and OvR approaches and to the perceptual assessment of SLPs.

\section{Automatic Classification Approach}
\label{sec: archs}
For the automatic classification of dysarthria, AoS, and neurotypical speech, we follow a hierarchical classification scheme with two classifiers as depicted in Fig.~\ref{fig: scheme}.
The first classifier SVM$_1$ is trained to discriminate between neurotypical speakers and patients (dysarthria or AoS) whereas the second classifier SVM$_2$ is trained to discriminate between dysarthria and AoS.
At test time, SVM$_1$ is first applied to decide whether the speaker is a neurotypical speaker or a patient.
If the speaker is classified to be a patient, SVM$_2$ is applied to decide whether the patient suffers from dysarthria or AoS.

Depending on the available speech material, speakers under consideration, and the classification objective (i.e., classifying neurotypical speakers and patients or classifying dysarthria and AoS), the discriminative power of different acoustic features can be different.
Hence, two feature selection blocks are incorporated.
The first block selects a subset of features that are optimal for classifying neurotypical speakers and patients and the selected features are used to train SVM$_1$.
The second block selects a subset of features that are optimal for classifying dysarthria and AoS and the selected features are used to train SVM$_2$.

In the following, additional details on the proposed approach are provided.
Further, OvO and OvR classification approaches considered to be automatic baseline approaches are briefly described.

\subsection{Acoustic features}

The acoustic features proposed in this paper for discriminating between dysarthria, AoS, and neurotypical speech are motivated by the advantageous performance these features have shown in discriminating between dysarthria and neurotypical speech in~\cite{Kodrasi_ITASLP_2019a} and in discriminating between dysarthria and AoS in~\cite{Kodrasi_Interspeech_2020}.
In the following, a brief overview of these features is presented.\footnote{For additional details on the motivation behind these features and their computation, the interested reader is referred to~\cite{Kodrasi_ITASLP_2019a, Kodrasi_Interspeech_2020}.}

\emph{Spectral sparsity.} \enspace
In~\cite{Kodrasi_ITASLP_2019a}, we have shown that spectral sparsity can successfully characterize imprecise articulation, abnormal pauses, and breathiness observed in dysarthria.
Spectral sparsity describes the energy distribution of the speech spectral coefficients across time and is computed by i) transforming the signals to the short-time Fourier transform domain, ii) time-aligning all representations to a reference representation, and iii) computing the shape parameter of a Chi distribution best modeling the spectral magnitudes in each time frame~\cite{Kodrasi_ITASLP_2019a}.
To manage the (possibly) high dimensionality of such a feature vector when the number of time frames is large, in this paper we do not time-align representations.
Instead, the shape parameter is computed for each time frame of the original representations and the used spectral sparsity feature vector $\mathbf{f}_{\text{1}}$ is a $4$--dimensional vector constructed by taking the statistics (i.e., mean, standard deviation, kurtosis, and skewness) of the so-computed shape parameter across all time frames.

\emph{Formant frequencies and duration of continuously voiced regions.} \enspace
As in~\cite{Kodrasi_Interspeech_2020}, vowel distortion and inappropriate vowel lengthening commonly observed in AoS are characterized by the $10$--dimensional feature vector $\mathbf{f}_2$ constructed from the statistics (i.e., mean, standard deviation, kurtosis, and skewness) of the first and second formant frequencies across time and the statistics (i.e., mean and standard deviation) of the duration of continuously voiced regions.

\emph{Loudness peaks per second and long-term average speech spectrum.} \enspace
As in~\cite{Kodrasi_Interspeech_2020, Hummel_Interspeech_2011, Berisha_ICASSP_2013}, abnormalities in loudness variation and hypernasality commonly observed in dysarthria are characterized by the \mbox{$10$--dimensional} feature vector $\mathbf{f}_{3}$ constructed by computing the number of loudness peaks per second and the mean speech power across time in nine octave bands.

\emph{Temporal sparsity.} \enspace
In~\cite{Kodrasi_ICASSP_2019, Kodrasi_ITASLP_2019a, Kodrasi_Interspeech_2020}, we have proposed to use temporal sparsity to characterize syllabification.
Temporal sparsity can be computed similarly as spectral sparsity, with the shape parameter modeling the speech spectral magnitudes in each frequency bin (rather than in each time frame).
Hence, to characterize syllabification, we construct the \mbox{$4$--dimensional} feature vector $\mathbf{f}_4$ by computing the statistics (i.e., mean, standard deviation, kurtosis, and skewness) of the shape parameter across all frequency bins.

\begin{figure}[t!]
  \centering
  \includegraphics[scale=0.7]{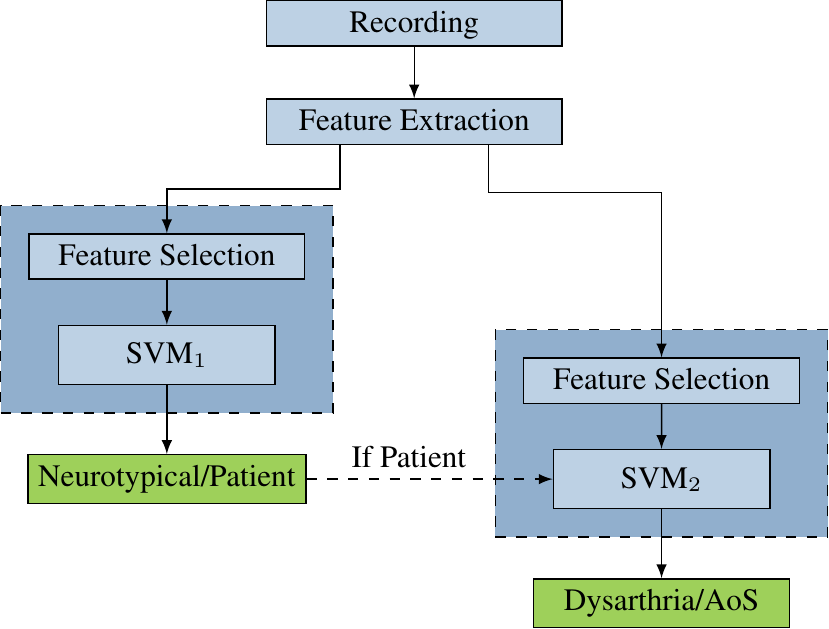}
  \caption{Schematic representation of the proposed approach for the automatic discrimination of dysarthria, AoS, and neurotypical speech. }
  \vspace{-0.5cm}
  \label{fig: scheme}
\end{figure}

Concatenating all previously described feature vectors into one vector, we obtain the $28$--dimensional feature vector $\mathbf{f}$ extracted in the feature extraction block in Fig.~\ref{fig: scheme}, i.e.,
\begin{equation}
\mathbf{f} = [\mathbf{f}^T_1, \; \mathbf{f}^T_2, \; \mathbf{f}^T_3, \; \mathbf{f}^T_4]^T.
\end{equation}

\subsection{Feature selection}
Although different feature selection approaches are applicable, statistical feature selection based on the ANOVA F-value is used in this paper~\cite{johnson_feature}.
Such an approach is independent of the used classifier and examines each feature individually.
Features are grouped according to the class label in the training data, the mean value of each feature across the different groups is analyzed, and a (user-defined) number of features showing the most significant differences between the different groups is selected.
The first feature selection block in Fig.~\ref{fig: scheme} selects optimal features for SVM$_1$ by analyzing whether there are significant differences in the mean values of the extracted features for neurotypical speakers and patients.
The second feature selection block selects optimal features for SVM$_2$ by analyzing whether there are significant differences in the mean values of the extracted features for patients with dysarthria and AoS.
The number of features selected by each block is determined based on the performance on the training set~(cf. Section~\ref{sec: auto_set}).


\subsection{Baseline automatic classification approaches} \enspace
As baseline classification approaches, we consider OvO and OvR classification. 
In the OvO approach, three SVMs are trained to discriminate between each class pair, i.e, neurotypical speech versus dysarthria, neurotypical speech versus AoS, and dysarthria versus AoS.
At test time, the class which receives most votes is selected for the final prediction.
In the OvR approach, three SVMs are trained to discriminate between each class and all other classes, i.e., neurotypical speech versus the rest, dysarthria versus the rest, and AoS versus the rest.
At test time, the class predicted with the most confidence (i.e., largest distance from the separating hyperplane) is selected for the final prediction.

\section{Automatic and Perceptual Classification}
\label{sec: results}
In this section, the performance of the proposed automatic hierarchical approach is compared to using the baseline OvO and OvR approaches.
To demonstrate the advantage of incorporating feature selection blocks, the performance when using all $28$ acoustic features in $\mathbf{f}$ (i.e., without feature selection blocks) is also investigated.
To demonstrate the advantage of using the proposed features $\mathbf{f}$, we also investigate the performance of all configurations (i.e., hierarchical with and without feature selection blocks, OvO, and OvR) using the openSMILE feature set from~\cite{Schuller_IS_2013}.
Finally, the automatic and perceptual classification accuracy are compared.

\subsection{Database and preprocessing}
\label{sec: preproc}
We consider French recordings collected at Geneva University Hospitals and University of Geneva of $29$ neurotypical speakers and $30$ patients, with $20$ patients diagnosed with dysarthria and the remaining $10$ patients diagnosed with AoS.
All patients with AoS have suffered a stroke, $14$ of the patients with dysarthria suffer from Parkinson's disease, and the remaining $6$ patients suffer from Amyotrophic Lateral Sclerosis.
There are $19$ female and $10$ male speakers in the neurotypical group and $12$ female ($6$ dysarthria, $6$ AoS) and $18$ male ($14$ dysarthria, $4$ AoS) speakers in the patient group.
The mean age of neurotypical speakers is $58.4$ years old and the mean age of patients is $66.0$ years old, with patients with dysarthria and AoS having a mean age of $72.8$ and $52.5$ years old, respectively.

The neurological diagnosis was established by neurologists, with the diagnosis of AoS based on the AoS rating scale~\cite{ars}.
The MSDs were assessed by an expert SLP using the perceptive score of BECD~\cite{BECD}. The BECD score reflects impairments in different dimensions such as voice quality, phonetic production, prosody, or intelligibility, and ranges from $0$ (no impairment) to $20$ (severe impairment).
The mean BECD score of all patients is $7.0$, where patients with dysarthria and AoS have a mean BECD score of $6.0$ and $9.1$, respectively. 



For the results presented in the following, we consider recordings of two sentences at a sampling frequency of $44.1$ kHz.
To ensure that the phonetic content does not influence classification results, all speakers in the database utter the same sentences.  
After downsampling to $16$ kHz and manually removing non-speech segments at the beginning and end of each sentence, the two sentences are concatenated and used to extract features (for automatic classification) or played back to judges (for perceptual classification).
The mean length of the concatenated sentences for the neurotypical speakers, patients with dysarthria, and patients with AoS is $15.1$~s, $16.4$~s, and $33.8$~s, respectively. 

\subsection{Performance evaluation}
\label{sec: perf_eval}

For automatic classification, the validation strategy is a stratified \mbox{$5$-fold} cross-validation ensuring that each fold has a balanced number of neurotypical speakers and patients and preserving the imbalanced distribution of patients with dysarthria and AoS
A single run of the $5$-fold cross-validation procedure may result in a noisy estimate of the automatic model performance, with different data splits possibly yielding different results.
For this reason, and in line with the number of perceptual evaluations available for each speaker (cf. Section~\ref{sec: auto_set}), we repeat the $5$-fold cross-validation procedure $10$ times such that a different random split of the data is used each time.
The performance is evaluated in terms of the mean and standard deviation of the balanced classification accuracy across all repetitions of the cross-validation procedure.
To compute the balanced classification accuracy, the individual classification accuracy for each group of speakers is first computed, i.e.,
\begin{equation}
  \label{eq: grp}
  \text{Acc}_{_{\text{Group}}} = \frac{\text{AP}_{\text{Group}}}{{\text{T}}_{\text{Group}}},
\end{equation}
with $\text{Group} \in \{ \text{Neurotypical, Dysarthria, AoS} \}$, $\text{AP}_{\text{Group}}$ denoting the number of accurately predicted speakers in the group, and $\text{T}_{\text{Group}}$ denoting the total number of speakers in the group.
The balanced classification accuracy is then defined as
\begin{equation}
  \text{Acc}_{_{\text{Balanced}}} = \frac{1}{3} \left( \text{Acc}_{_{\text{Neurotypical}}} + \text{Acc}_{_{\text{Dysarthria}}} + \text{Acc}_{_{\text{AoS}}}\right).
\end{equation}
In addition to the individual and balanced classification accuracy, the classification accuracy for all patients $\text{Acc}_{_{\text{Patient}}}$ is considered in Section~\ref{sec: res}.
$\text{Acc}_{_{\text{Patient}}}$ is defined as in~(\ref{eq: grp}), with $\text{AP}_{_{\text{Patient}}}$ being the number of patients with dysarthria and AoS that are accurately predicted to be patients (independently of whether the exact label, i.e., dysarthria or AoS, is correct) and $\text{T}_{_{\text{Patient}}} = \text{T}_{_{\text{Dysarthria}}} + \text{T}_{_{\text{AoS}}}$.

\subsection{Automatic and perceptual classification settings} \enspace
\label{sec: auto_set}

\emph{Automatic classification.} \enspace For automatic classification, we use SVMs with a radial basis kernel function.
To select the soft margin constant $C$  and the kernel width $\gamma$ for the SVMs, nested $5$-fold cross-validation is performed on the training data in each fold, with $C \in \{10^{-2}, 10^{4} \}$ and $\gamma \in \{10^{-4}, 10^{2} \}$.
To set the number of features $n_f$ that the feature selection blocks should select, nested $5$-fold cross-validation is performed on the training data in each fold, with $n_f \in \{5, 10, 15, 20\}$.
The final hyper-parameters (i.e., $C$, $\gamma$, and $n_f$) used in each fold are selected as the ones resulting in the highest mean balanced accuracy on the training data.

\emph{Perceptual classification.} \enspace
For perceptual classification, $20$ SLPs were recruited as judges.
The judges were French native speakers and had on average $11$ years of professional experience.
The perceptual classification task was done following a similar methodology as for the automatic classification scheme in Fig.~\ref{fig: scheme}, i.e., judges listened to the available recordings and for each recording they were asked to decide: i) whether the recording belonged to a neurotypical speaker or a patient and ii) if the recording belonged to a patient, whether the patient suffered from dysarthria or AoS.
To minimize the duration of the perceptual task for each judge, we split the available recordings into two groups, with one group containing the recordings of $15$ neurotypical speakers, $10$ patients with dysarthria, and $5$ patients with AoS and the other group containing the remainder of the recordings (i.e., $14$ neurotypical, $10$ dysarthria, and $5$ AoS).
Consequently, $10$ judges were asked to evaluate recordings belonging to one group and $10$ judges were asked to evaluate recordings belonging to the other group.
Hence, in line with the automatic classification results where each recording was evaluated by $10$ different automatic models through repetitions of the cross-validation procedure, each recording was also perceptually evaluated by $10$ different judges.
The perceptual classification performance is then computed as described in Section~\ref{sec: perf_eval}, with the mean and standard deviation of the performance computed across judges.

\begin{table}[t]
  \footnotesize
  \begin{center}
    \caption{\footnotesize Mean and standard deviation of the balanced classification accuracy $\text{Acc}_{_{\text{Balanced}}}$ [\%] using several configurations: the proposed hierarchical approach with feature selection blocks, the hierarchical approach without feature selection blocks, and the baseline OvO and OvR approaches.
    The performance of all these configurations using the proposed handcrafted feature set $\mathbf{f}$ and the openSMILE feature set is presented.}
    \label{tbl: perf_auto}
    \def\tabcolsep{2.5pt}
    \begin{tabularx}{\linewidth}{Xrr}
      \toprule
      Classification approach & Handcrafted $\mathbf{f}$ & openSMILE \\
      \toprule
      Proposed (hierarchical with feature selection) & $ 79.7 \pm 4.0$ & $74.4 \pm 3.4$ \\
      Hierarchical without feature selection &  $75.0 \pm 3.4 $ & $58.8 \pm 3.9$ \\
      OvO & $72.0 \pm 3.8$  & $57.3 \pm 4.0$ \\
      OvR & $74.8 \pm 2.4$  & $60.8 \pm 4.4$\\ \hline
     \bottomrule 
   \end{tabularx}
   \vspace{-0.5cm}
 \end{center}
\end{table}

\subsection{Results}
\label{sec: res}

Table~\ref{tbl: perf_auto} presents the balanced classification accuracy obtained using all considered configurations with the handcrafted feature set $\mathbf{f}$ and the openSMILE feature set.
It can be observed that the proposed classification approach using the handcrafted features $\mathbf{f}$ achieves the best performance, with the incorporation of feature selection blocks increasing the balanced classification accuracy from $75.0$\% to $79.7$\%.
These results confirm that different subsets of features are optimal for different classifiers.
The performance difference when incorporating feature selection blocks is even larger for the openSMILE feature set, since this is a high-dimensional feature vector (i.e., $6373$) overfitting to the training data when feature selection is not used. 
Further, the presented results show the advantages of using the proposed hierarchical classification approach rather the OvO and OvR approaches.\footnote{It should be noted that the proposed approach also outperforms using the OvO and OvR approaches with additional feature selection blocks. However, due to space constraints, these results are not presented here.}
In the following, the performance of the proposed hierarchical classification approach (using the handcrafted features $\mathbf{f}$ and feature selection blocks) is further analyzed and compared to the perceptual performance achieved by SLPs.

Table~\ref{tbl: perf_auto_perc} presents the automatic and perceptual classification performance.
It can be observed that automatic classification yields a higher performance than perceptual classification in terms of all considered accuracy measures.
The difference in accuracy is particularly large for neurotypical speakers and patients with dysarthria.
While the classification accuracy for neurotypical speakers is $82.1$\% using automatic classification, the perceptual accuracy is only $67.2$\%.
Further, while the classification accuracy for patients with dysarthria is $75.0$\% using automatic classification, the perceptual accuracy is only $64.5$\%.
Since the impairment for patients with dysarthria can be milder than for patients with AoS (as shown by the lower mean BECD score presented in Section~\ref{sec: preproc}), judges often confuse neurotypical speakers for patients with dysarthria and conversely.
Although the classification accuracy for neurotypical speakers is lower than for patients also for automatic classification~(i.e., $82.1$\% versus $91.7$\%), confusion between neurotypical speakers and patients with dysarthria does not occur as often in the proposed automatic classification scheme.

\begin{table}[t!]
  \begin{center}
    \caption{\footnotesize Mean and standard deviation of the classification accuracy for different groups of speakers using automatic and perceptual classification. Automatic classification is done using the hierarchical approach proposed in Section~\ref{sec: archs}. Perceptual classification is done by SLPs as described in Section~\ref{sec: auto_set}.}
    \label{tbl: perf_auto_perc}
    \def\tabcolsep{2pt}
    \begin{tabularx}{\linewidth}{X|rr}
      \toprule
      Accuracy [\%] &  Automatic classification & Perceptual classification \\
      \toprule
      $\text{Acc}_{_{\text{Balanced}}}$ & $79.7 \pm 4.0$ & $68.9 \pm 4.3\color{white}{0}$ \\
      \hline
      $\text{Acc}_{_\text{Neurotypical}}$ & $82.1 \pm 1.4 $ & $67.2 \pm 9.2\color{white}{0}$ \\
      $\text{Acc}_{_\text{Patient}}$ & $91.7 \pm 3.7$ & $82.7 \pm 7.0\color{white}{0}$ \\
      $\text{Acc}_{_\text{Dysarthria}}$ &  $75.0 \pm 7.7 $ & $ 64.5 \pm 10.1$ \\
      $\text{Acc}_{_\text{AoS}}$ &  $82.0 \pm 7.5 $ & $75.0 \pm 14.3$ \\ \hline
     \bottomrule 
   \end{tabularx}
   \vspace{-0.5cm}
 \end{center}
\end{table}

In addition, the presented results show that the trend in performance for individual groups of speakers is similar for both automatic and perceptual classification approaches, i.e., $\text{A}_{_{\text{Patient}}} \!\!>\!\! \text{A}_{_{\text{Neurotypical}}}$ and $\text{A}_{_{\text{AoS}}} \!\!>\!\! \text{A}_{_{\text{Dysarthria}}}$.
This similar trend is to be expected since the acoustic features used in the proposed automatic classification technique are motivated by the clinical-perceptual signs used to diagnose these MSDs and since the automatic and perceptual classification approaches follow a similar hierarchical methodology.

In summary, the presented results show that the proposed automatic classification scheme can be an advantageous tool to integrate in clinical practice. 
In addition, the presented results show that while automatic tools can achieve a high performance in discriminating patients from neurotypical speakers, the performance in discriminating subtypes of MSDs needs to be improved.
Analyzing the generalisability of the presented results to other databases and analyzing the statistical significance of the performance differences between automatic and perceptual classification remain topics for future investigation.

\vspace{-0.25cm}

\section{Conclusion}
In this paper, we have proposed a hierarchical three-class automatic technique operating on handcrafted acoustic features for the discrimination of dysarthria, AoS, and neurotypical speech.
Two SVMs are used, with the first SVM discriminating between impaired and neurotypical speech and the second SVM discriminating between dysarthria and AoS.
Since the discriminative power of different features is expected to be different for different groups of speakers, two feature selection blocks are incorporated prior to the two SVMs.
The classification accuracy of this approach has been analyzed on a French database of dysarthria, AoS, and neurotypical speech.
Additionally, a perceptual study has been conducted where SLPs are asked to discriminate between dysarthria, AoS, and neurotypical speech on the same database.
The presented results have shown the advantages of the automatic classification technique, which yields a balanced classification accuracy of $79.7$\% in comparison to the balanced accuracy of $68.9$\% achieved in the perceptual assessment of SLPs.

\footnotesize
\bibliographystyle{IEEEbib}
\bibliography{refs}

\end{document}